\documentclass[]{amsart}
\usepackage[utf8]{inputenc}
\usepackage{hyperref}
\usepackage{url}
\usepackage{multirow}
\usepackage{makecell}
\usepackage{color}
\usepackage{adjustbox}
\usepackage{graphicx}
\usepackage{cite}
\usepackage{balance}
\usepackage{pifont}
\usepackage{tabularx}
\usepackage{array}
\usepackage{makecell}
\usepackage{comment}
\usepackage{soul}
\usepackage{xfrac}
\usepackage{amsmath}
\usepackage{xcolor}
\usepackage{amsfonts}
\usepackage{listings}
\usepackage{amssymb}
\usepackage{amsmath}
\usepackage{authblk}
\usepackage{multicol}

\usepackage[margin=3cm]{geometry}

\DeclareRobustCommand{\gobblefive}[5]{}
\newcommand*{\SkipTocEntry}{\addtocontents{toc}{\gobblefive}}

\RequirePackage{caption}
\RequirePackage{subcaption}

\definecolor{white}{RGB}{255,255,255}    
\definecolor{greyYed}{RGB}{204,204,204}  
\definecolor{yellowYed}{RGB}{255,204,0}  
\definecolor{greenYed}{RGB}{153,204,0}   
\definecolor{orangeM}{RGB}{217, 83, 25}  
\definecolor{blueM}{RGB}{0, 114, 189}    
\definecolor{yellowM}{RGB}{237, 177, 32} 
\definecolor{purpleM}{RGB}{126, 47, 142} 
\definecolor{greenM}{RGB}{119, 172, 48}  
\definecolor{cyanM}{RGB}{77, 190, 238}   
\definecolor{redM}{RGB}{162, 20, 47}     

\newcolumntype{C}[1]{>{\centering\arraybackslash}p{#1}}

\newcolumntype{L}[1]{>{\arraybackslash}p{#1}}

\newcommand{\ie}{\textit{i.e.}\ }

\newcommand{\NN}{\mathbb{N}}

\newcommand{\doi}[1]{[\textsc{doi:}\href{https://www.doi.org/#1}{#1}]}

\makeatletter
\def\blfootnote{\gdef\@thefnmark{}\@footnotetext}
\makeatother

\title[
    Characterizing Polkadot's Transactions Ecosystem
]{
    Characterizing Polkadot's Transactions Ecosystem: methodology, tools, and insights
}

\author[Maurantonio Caprolu, Roberto Di Pietro, Flavio Lombardi, and Elia Onofri]{}

\date{}

\begin{document}

\blfootnote{\phantom{x}}

\maketitle

\vspace{-1em}

\begin{center}
    \begin{minipage}{.49\linewidth}\centering
        \textsc{Maurantonio Caprolu}\\
        \medskip
        {\footnotesize
            King Abdullah University\\
            of Science and Technology (KAUST),\\
            Thuwal 23955, Saudi Arabia\\
            \texttt{rdipietro@kaust.edu.sa}
        }
    \end{minipage}
    \begin{minipage}{.49\linewidth}\centering
        \textsc{Roberto Di Pietro}\\
        \medskip
        {\footnotesize
            King Abdullah University\\
            of Science and Technology (KAUST),\\
            Thuwal 23955, Saudi Arabia\\
            \texttt{mcaprolu@kaust.edu.sa}
        }
    \end{minipage}
    
    \bigskip\bigskip
    
    \begin{minipage}{.49\linewidth}\centering
        \textsc{Flavio Lombardi}\\
        \medskip
        {\footnotesize
            Istituto per le Applicazioni del Calcolo,\\
            Consiglio Nazionale delle Ricerche (IAC--CNR),\\
            Rome 00185, Italy\\
            \texttt{flavio.lombardi@cnr.it}
        }
    \end{minipage}
    \begin{minipage}{.49\linewidth}\centering
        \textsc{Elia Onofri}\\
        \medskip
        {\footnotesize
            Istituto per le Applicazioni del Calcolo,\\
            Consiglio Nazionale delle Ricerche (IAC--CNR),\\
            Rome 00185, Italy\\
            \texttt{onofri@iac.cnr.it}
        }
    \end{minipage}
\end{center}

\medskip
\thispagestyle{empty}

\begin{abstract}
The growth potential of a crypto project, typically sustained by an associated cryptocurrency, can be measured by the use cases spurred by the underlying technology. However, these projects are usually distributed, with a weak (if any) feedback schemes. 
Hence, a metric that is widely used as a proxy for the healthiness of such projects is the number of transactions and related volumes. Nevertheless, such a metric can be subject to manipulation---the crypto market being an unregulated one, magnifies such a risk. To address the cited gap, in this paper, we design a comprehensive methodology to process large cryptocurrency transaction graphs that, after clustering user addresses of interest, derives a compact representation of the network that highlights interactions among clusters. It is the analysis of these interactions that provides insights into the strength of the project.

To show the quality and viability of our solution, we bring forward a use case centered on Polkadot.
The Polkadot network, a cutting-edge cryptocurrency platform, has gained significant attention in the digital currency landscape due to its pioneering approach to interoperability and scalability. However, little is known about how many and to what extent its wide range of enabled use cases have been adopted by end-users so far. The answer to this type of question means mapping Polkadot (or any analyzed crypto project) on a palette that ranges from a thriving ecosystem to a speculative coin without compelling use cases. 

Our findings, rooted on extensive experimental results---we have parsed 12.5+ million blocks---, demonstrate that crypto exchanges exert considerable influence on the Polkadot network, 
owning nearly 40\% of all addresses in the ledger and
absorbing at least 80\% of all transactions. In addition, the high volume of inter-exchange transactions (more than 20\%) underscores the strong interconnections among just a couple of prominent exchanges, prompting further investigations into the behavior of these actors to uncover potential unethical activities, such as wash trading. \\*
These results are a testament to the quality and viability of the proposed solution that, while characterized by a high level of scalability and adaptability, is at the same time immune from the drawbacks of currently used metrics.

\medskip


\end{abstract}

\begin{center}
    \textsc{Contents}
\end{center}

\begin{multicols}{2}
    \tableofcontents
    \vspace{-2em}
\end{multicols}

\newpage

\section{Introduction}\label{sec:intro}

The cryptocurrency market has emerged as a dynamic and rapidly evolving sector in the global financial landscape, with a capitalization in the order of Trillions USD. 
Cryptocurrencies offer several advantages over traditional fiat currencies, such as decentralization, peer-to-peer transactions without intermediaries, increased financial inclusion, and the potential for significant returns on investment. However, alongside these advantages, cryptocurrencies also suffer from a few drawbacks and are subject to idiosyncratic risks~\cite{8892660}. One notable concern is the high volatility of the cryptocurrency market, which can lead to substantial price fluctuations over short periods. Also, regulatory uncertainties and the lack of protection mechanisms for investors pose challenges to market participants. Furthermore, the pseudo-anonymity provided by cryptocurrencies can facilitate illicit activities, mainly related to money laundering, market manipulation, and other frauds~\cite{dupuis2020money}.

At the time of writing this paper, the cryptocurrency market consists of more than 2 million different coins, with a total market capitalization of more than 2.5 trillion\footnote{Data taken from \url{https://coinmarketcap.com/} on March $17^{th}$ 2024.}. 
This impressive size increases the challenges of navigating the cryptocurrency landscape and capitalizing on its opportunities, particularly given the prevalence of scam coins.
In fact, numerous questionable projects promise high returns or revolutionary technologies, but fail to deliver on their promises, resulting in substantial financial losses for investors~\cite{9591634}. These scams often use different techniques, such as fake ICOs (Initial Coin Offerings) and Ponzi schemes, to deceive unsuspecting investors~\cite{li2023double}.
Identifying scam coins can be relatively straightforward for practitioners with expertise in financial securities, who can resort to several techniques, such as investigating the project's team members, the technologies used, and the published white papers and roadmaps, to assess the credibility of a particular project.
However, evaluating the potential for success of well-known and reputable projects is considerably more complex, even for seasoned crypto investors. In fact, even if a project has robust underlying technologies, a truly decentralized network, and an experienced developer team behind it, its success is not always guaranteed. 
One of the most important ingredients is a strong and engaged community, which usually contributes significantly to the success of a cryptocurrency. Generally, a large and active user base promotes adoption, fosters network effects, and supports development and ecosystem growth. In addition, successful cryptocurrencies usually offer practical utility, serve specific use cases, or solve real-world problems. Whether it is facilitating peer-to-peer transactions, enabling decentralized finance (DeFi) applications, or providing solutions for identity management or supply chain tracking, cryptocurrencies with tangible use cases tend to gain traction. 

In contrast, a cryptocurrency that lacks a specific use case or fails to address a significant problem may only be used as a speculative medium~\cite{DiPietro2021}. In such cases, its value is primarily driven by trading activity and market sentiment rather than intrinsic utility, making the cryptocurrency vulnerable to substantial price fluctuations. Then, when the trading community perceives the coin as unprofitable, interest in the project may decrease, leading to a drop in trading volume and liquidity and an inevitable decline in market capitalization. 

Hence, one of the best metrics to assess a project's health status and growth potential is to evaluate active use cases.
However, this metric is very challenging to obtain, since the project can be (and usually is) a distributed one, with weak to nil reporting. That is why metrics that could act as proxy have been sought. One of the most diffused one is the cryptocurrency's transactions number and volume.
The rational behind it is that, if the cryptocurrency shows healthy trade level (with respect to the market cap), that implies that there is an active interest in the underlying project.
And that is exactly where Goodhart's Law\footnote{Like almost all economics theories it is not really a law though, unlikely almost all economics theories, it seems supported by empirical observations~\cite{GoodhartLaw}.} comes into play. In its popular form, it states that: ``when a measure becomes a target, it ceases to be a good measure.''.
Indeed, the cited metric has two critical weaknesses: (i) an ontological one---simply examining transactions number and volume does not provide insight into how users interact with the network; and, (ii) a systemic one---these statistics may be altered by some unethical phenomena, such as wash trading, allegedly performed by crypto exchanges~\cite{CHEN2022126405,PENNEC2021101982}, the largest players in the cryptocurrency market.

Centralized crypto exchanges, also known as ``custodial'' exchanges, are usually the first entry point into any cryptocurrency network. They offer an easy-to-use platform to buy digital assets and trade among supported cryptocurrencies. Their user-friendly interface and effective technical support make centralized exchanges the preferred trading platform for all cryptocurrencies in the market, in contrast to decentralized exchanges, which are still relatively complex and hard to use for neophytes.
Given their peculiar role as intermediaries linking the virtual realm of cryptocurrencies with the tangible world of fiat currencies, investigating the on-chain activities of crypto exchanges can unveil valuable insights into the speculative dynamics surrounding a cryptocurrency. Indeed, users interact on-chain with crypto exchanges to deposit and withdraw crypto coins. Therefore, if a user's transactions are primarily limited to exchanges rather than with other entities, it is reasonable to categorize them as traders---their objective being to make a profit, while not being interested in the finalities of the project underlying the speculated crypto asset. 

In this paper, we leverage the above introduced observation to investigate the ecosystem induced by centralized crypto exchanges within the Polkadot network.

Polkadot, a cutting-edge cryptocurrency platform, has gained significant attention in the digital currency landscape thanks to its innovative approach to interoperability and scalability. Its robust technology, coupled with a vibrant and engaged community, places Polkadot as a frontrunner in the race to revolutionize the blockchain industry. Theoretically, Polkadot has all the elements to be a successful project. Indeed, investors quickly recognized its potential, propelling the project to consistently rank among the top 15 cryptocurrencies by market capitalization. However, little is known about how users interact with the platform, and how much traction for its wide range of use cases---spanning DeFi, decentralized applications, cross-chain communication, governance, and more---, currently measured through the proxy metric of transactions, is real.  

{\bf Contribution.}
The main goal of this paper is to gain a deeper understanding of the contribution of crypto exchanges to the transactions related to the Polkadot network and, by extension, to evaluate the network's exposure to speculative activities.
To this end, we introduce an innovative methodology designed to cluster addresses associated with crypto exchanges and, by applying a contraction algorithm, derive a compact representation of the transaction network to better visualize the ecosystem fostered by crypto exchanges.
The main contributions of this work can be summarized as follows:
\begin{itemize}
    \item We present a novel methodology able to process large cryptocurrency transaction graphs to investigate the interactions among different categories of users; the methodology starts with the building of the transaction graph and includes a clustering strategy and a contraction algorithm.
    
    \item Our methodology enables: (i) a thorough understanding on the interactions between different categories of users, leading to a robust assessment of how much a cryptocurrency network is used for speculative purposes; and, (ii) a characterization, 
    thanks to our contracted transaction graph, of the behavior of crypto exchanges, potentially leading to uncovering unethical activities, such as wash trading.
    
    \item We apply our methodology to the Polkadot network, one of the most successful and promising projects in the market, to investigate the influence exerted by crypto exchanges on the transaction network. Our experimental analysis involved collecting and analyzing more than 12.5 million blocks---from genesis up to block 12,532,600.
    
    \item We show several novel findings on the interaction between different categories of Polkadot users. Among the other insights, we detect a huge number of transactions involving crypto exchanges (around 80\%), suggesting that Polkadot users are there mainly for speculation rather than for contributing to the supported use cases.
    
    \item We provide a preliminary analysis of the exchanges' on-chain activities, including inter-exchanges connections, highlighting strong relations among a few well-known exchanges, and high volumes of intra-exchanges transactions.
    
    \item We release, under open source license, the code base and our two graphs, i.e., both original and contracted versions, to ensure the reproducibility of our results and allow further investigations in the same area\footnote{A link to our repository will be included in the final (non-preprint) version of this manuscript.}.
\end{itemize}

Although tailored for the Polkadot network, our methodology is of general applicability, and can work directly to any account-based cryptocurrency. With minor adjustments, it can also be adapted to UTXO-based coins. In addition, our methodology features a high degree of adaptability; for instance, it can integrate additional user categories beyond crypto exchanges, enriching the compact graph and enabling flexible adaptation to different analyses.

\section{Related Work} \label{sec:related}
Transaction graph analysis is a fundamental tool in the study of cryptocurrencies, which can provide valuable insights into various aspects of the cryptocurrency ecosystem. A transaction graph allows the study of money flows, playing a crucial role in understanding the status of a network, the user behavior, and detecting fraud and illicit activities. For this reason, several works leveraged graph analysis to investigate cryptocurrencies, with Bitcoin and Ethereum as preferred targets~\cite{serena2021cryptocurrencies}. Among its wide range of use cases, graph analysis has been used in the context of cryptocurrencies to uncover different properties of the Bitcoin~\cite{maesa2019bow} and Ethereum~\cite{10.1145/3381036} networks, and many different tasks, such as price prediction~\cite{8964468}, address clustering and other attacks to pseudo-anonymity~\cite{Harrigan2016,10.1007/978-3-319-70278-0_9}, deanonymization~\cite{JAWAHERI2020101684}, and anomaly detection~\cite{arijit2022graph}.

Although highly successful in the cryptocurrency market, Polkadot has not received the same level of attention from academia. However, despite the Polkadot architecture is described and discussed mainly in the gray literature, a few recent research works have analyzed its network. The Polkadot architecture was detailed in~\cite{abbas2022analysis}, which identified several limitations and contradictions after conducting a data-driven study on the transaction network. Polkadot user transactions have been formally modeled and analyzed in~\cite{Hanaa_FC_24}, where the authors presented numerous insights into the network topology using graph analysis. Finally,~\cite{10174938} presented a deposit address detection strategy used to cluster crypto exchange's customers in an attempt to reduce the pseudo-anonymity ensured by the Polkadot protocol. 

To the best of our knowledge, this is the first study investigating the ecosystem generated by crypto exchanges on the Polkadot network through graph analysis.

\section{Methodology}\label{sec:methodology}
Our methodology includes multiple steps. First, we build the Polkadot transaction graph following the network modeling proposed in~\cite{Hanaa_FC_24}. Then, we identify crypto exchanges among the most central nodes in the transaction graph by leveraging the ``deposit address reuse'' heuristic presented in~\cite{10174938}.
Finally, we obtain a compact representation of the Polkadot transaction graph by applying the contraction algorithm proposed in~\cite{LombardiOnofri22a} with the list of exchanges discovered in the previous step as input. 
The steps mentioned above are discussed in more detail in the following sections.

\subsection{Polkadot Transaction Graph}\label{subsec:tx_graph}
In the Polkadot ecosystem, each account has a unique address that serves as its identifier in the network. 
Whenever a transaction occurs, it involves transferring a certain amount of the native cryptocurrency, called \emph{DOT}, between a sender and a recipient, which are Polkadot addresses. These addresses become nodes in the graph, and the transaction between them becomes an edge. 
Similarly to Ethereum, the Polkadot transaction system can be modeled as a multigraph 
$G = (V, E)$, where $V$ is the set of Polkadot addresses and $E$ is the set of all transactions included in the ledger~\cite{Hanaa_FC_24}. 
Such graph is weighted, where weights are attributes that describe the edges of the graph. Attributes include the transaction value, expressed in \emph{DOT}, and the transaction timestamp. It is important to note that a multigraph allows an arbitrary number of edges to exist between a pair of nodes in any direction, reflecting the account-based nature of the Polkadot network.

\subsection{Exchanges Detection}\label{subsec:exch_detection}
Many strategies can be used to detect centralized crypto exchanges on a cryptocurrency transaction graph. 
First, some crypto exchanges publicly disclose their deposit addresses for various cryptocurrencies on their websites or through other communication channels. However, this information is usually limited to the main wallet address---a crypto exchange can use many. Moreover, it is generally difficult to collect and validate such information, if found in unofficial sources. \\*
Another possibility is pattern analysis in the transaction graph. In fact, addresses belonging to exchanges tend to receive a high volume of transactions, as users regularly deposit and withdraw funds into their exchange platform for trading purposes. Analyzing the volume of transactions, together with their values, can help identify addresses that belong to exchanges. 
Nevertheless, when using transaction volume only, it is difficult to identify a pattern that is exclusively used by crypto exchanges. Therefore, this methodology can easily lead to a high false positive rate. Furthermore, this detection strategy works better with larger exchanges, i.e., with a considerable number of customers, neglecting smaller ones, which may go unnoticed. \\*
A more robust detection strategy is based on a distinctive feature inherent in decentralized exchanges: the deposit address. When users want to deposit crypto tokens into their exchange wallet, they are provided with a specific deposit address designated to receive their funds. This address, generated and owned by the exchange, is unique and unchanged for each user, ensuring simplicity and avoiding unnecessary burden on the cryptocurrency network. 
As a consequence, the transactions from a user-owned address to a deposit address, and the subsequent one from the deposit address to the exchange main wallet, generate a clear pattern easily identifiable and valuable for detecting exchange's main addresses. This strategy has previously been used to cluster crypto exchange customers on the Ethereum~\cite{Friedhelm_FC_2020} and Polkadot~\cite{10174938} networks.

In this work, one of our main goals is to cluster Polkadot accounts owned by crypto exchanges, including their main wallets and deposit addresses. Leveraging the Polkadot transaction graph generated in the preceding step, we focus on analyzing nodes with the highest centrality in terms of transaction volume, as they are most likely to be linked to crypto exchanges. In particular, for each of these central nodes, we test all its neighbors using the deposit address detection methodology outlined in~\cite{10174938}. Subsequently, any node that exhibits more than 90\% of deposit addresses among its neighbors is classified as a crypto exchange. Finally, we put in the same cluster the analyzed node---as the exchange's main wallet---and all its neighbors identified as deposit addresses. This clustering forms the basis for the contraction algorithm discussed in the following section.

\subsection{Contraction Algorithm}\label{subsec:contraction}
Given a large graph $G = (V, E)$, it is common in real applications to have some sort of typological information that categorizes the vertices.
Such pieces of information might come from intrinsic data on the graph, or they can be derived from some external expert knowledge on the subject of the analysis.
Formally, such categorization can be seen as a (non-proper) coloring function $\gamma : V \to C$, where $C \subset \NN$ is a countable set of colors.

Given $G$ and $\gamma$, we can then consider the clustering obtained by regrouping all the connected vertices sharing the same color, hence inducing a color partition of the vertices.
Graph contraction is a natural way to retrieve a compact representation of such partition, progressively merging all the couples of connected nodes sharing the same colors.
Such an operation---we will be  referring to  with the name of $\gamma$-contraction (or color-contraction on the coloring $\gamma$)---is independent with respect to the order of the application of said contraction operations, 
as it was widely discussed in \cite{Onofri23}.

Formally, a $\gamma$-contraction yields a contracted graph $G/\gamma = G' = (V', E')$, where each node $v' \in V'$ represents a maximal set of nodes $S \subset V$ such that $\gamma(v') = \gamma(v), \forall v \in S$ and where the spanning graph $\langle S\rangle$ induced by $S$ over $G$ is connected.
$v'$ then agglomerates the data related to $S$ in a single vertex that can be equipped with useful pieces of information including, but not limited to, 
the order and the size of $\langle S \rangle$.
Analogously, $e' \in E'$ joins two vertices $u', v' \in V'$ obtained from two sets of nodes $R, S \subset V$ if $\exists u \in R, v \in S$ such that $(u, v) \in E$.
The semantic of $e'$ is that it aggregates the data related to all the edges insisting between $R$ and $S$ into a single edge that can be enriched with the cardinality of such connections---along with any other useful piece of information.

In particular, it follows from the maximality of $R$ and $S$ that $\gamma(u') \ne \gamma(v') \forall (u', v') \in E'$, hence $\gamma$ is a proper coloring of $G'$. This ensures that the so formed representation retrieves the independent sets of nodes with respect to colors, hence highlighting the interactions between the different clusters.

\section{Experimental Setup}\label{sec:experimental}
Our experimental campaign involves multiple steps. 
We start with data collection, where we download the ledger from the Polkadot network. 
Then, we build the transaction graph, cluster the addresses that belong to exchanges, and contract the transaction graph. 
Finally, we analyze the contracted graph to gain insights into the crypto exchange ecosystem in the Polkadot network. 

\subsection{Data Collection}

The first step of our methodology is to retrieve data from the Polkadot ledger. To this end, we install a Polkadot full node on a workstation running Ubuntu 20.04 LTS. 
Then, we set up the full node in archive mode to access and download the full ledger. In the Polkadot blockchain, implemented over the Substrate framework, a block is made up of a header and a body. The body contains a list of objects, called {\em extrinsics}. An extrinsic is a generalization of the concept of transaction that represents changes in the state of the network coming from the outside world. An extrinsic can contain many different types of data that need to be validated by the community and stored in the ledger, such as fund transfers, staking rewards, and council proposals, to name a few~\cite{abbas2022analysis}.

After fully syncing the blockchain ledger on our full node, we retrieved and parsed the blocks data from the genesis (mined in May 26, 2020), up to October 18, 2022, for a total of more than 12.5 million blocks. 
We queried our Polkadot full-node using two open source Python libraries implemented by Parity Technologies\footnote{\url{https://www.parity.io/about}}. In particular, we interacted with the Polkadot ledger through \texttt{substrate-interface}, which allowed us to retrieve row data related to blocks and extrinsics. 
Then, we decoded transaction data, encoded in SCALE (Simple Concatenated Aggregate Little-Endian) format, with \texttt{scalecodec}.
Since we are only interested in user-initiated fund transfers that were successfully validated and stored in the ledger, we have collected only extrinsics that have the attributes reported in Figure~\ref{fig:extrinsic}. 

\begin{figure}
    \centering
    \begin{minipage}{.7\linewidth}
        \begin{lstlisting}[
            language=Python,
            showspaces=false,
            basicstyle=\footnotesize\ttfamily,
            frame=bt,
            commentstyle=\color{gray}
        ]
module_id = Balances; and
call_id = Transfer or transfer_keep_alive or 
transfer_all; and
signed = True; and
Success = 1
        \end{lstlisting}
    \end{minipage}
    \caption{List of attributes that identify an extrinsic as a successfully completed DOTs transfer between users, i.e., validated by the community and stored in the ledger.}
    \label{fig:extrinsic}
\end{figure}

In this way, we retrieved and stored all extrinsics associated with the successful transfers of \emph{DOT}s between users where the transaction amount was debited from the sender's account and deposited to the recipient's account. 

\begin{table}[]
    \renewcommand{\arraystretch}{1.3}
    \centering   
    \begin{tabular}{cc||c|c}
         & & \textbf{$G$} & \textbf{$G/\gamma$}\\\hline\hline
        \multirow{6}{*}{\textbf{Accounts}} & \multirow{2}{*}{\textbf{total}} & \multirow{2}{*}{2,261,478} & 654,446 \\
         & & & $(28.94\%)$ \\\cline{2-4}
         & \multirow{2}{*}{\textbf{exchanges}} & 877,956 & \multirow{2}{*}{33} \\
         & & $(38.82\%)$ & \\\cline{2-4}
         & \multirow{2}{*}{\textbf{users}} & 1,383,522 & \multirow{2}{*}{654,413} \\
         & & $(61.18\%)$ & \\\hline
        \multirow{4}{*}{\textbf{Transactions}} & \multirow{2}{*}{\textbf{number}} & \multirow{2}{*}{8,055,559} & 942,761 \\
         & & & $(11.74\%)$ \\\cline{2-4}
         & \multirow{2}{*}{\textbf{total flux}} & \multicolumn{2}{c}{\multirow{2}{*}{9,197,251,270.61}}\\
         & & \multicolumn{2}{c}{}
    \end{tabular}
    \caption{Network statistics}
    \label{tab:networkStatistics}
\end{table}
\subsection{Transaction Graph and Exchanges Detection}
After downloading the Polkadot ledger, we used the Python library NetworkX to build a \textit{MultiDiGraph} $\bar G = (V, \bar E)$ representing the Polkadot account-to-account transaction network, as discussed in Section~\ref{subsec:tx_graph}.
Although data collection began with the genesis block, our analysis starts at block 1,205,128, since the balance transfer functionality was enabled following this block height.
The dimensions of the resulting multigraph $\bar G$ are shown in the first column of Table~\ref{tab:networkStatistics} (where $G$ is a pre-processed version of $\bar G$ sharing the same topological statistics with it---details in Section~\ref{subsec:contraction-setup}). The network size accounts for more than 2.2 million accounts, that generated more than 8 million transactions, moving a total of 9 trillion \emph{DOT}s. 

After building the Polkadot transaction graph, we computed the degree distribution to gather useful information on how nodes are connected to each other. In particular, we measured the degree centrality to identify the most important nodes in the network, i.e., with the highest number of incoming and outgoing transactions. Then, we analyzed the 60 most central nodes to investigate whether they behave like a crypto exchange or not. In particular, for each node under analysis, we check if its neighbors, i.e., other nodes at distance 1, are deposit addresses by using the methodology discussed in Section~\ref{subsec:exch_detection}. We found that 41 of the 60 most central nodes on the Polkadot transaction graph belong to 33 different crypto exchanges.
In total, the 33 exchanges owned more than 800k Polkadot addresses, accounting for 38.82\% of all addresses in the network, as shown in Table~\ref{tab:networkStatistics}. We stopped our analysis after examining the 60th most central node. In fact, beyond the 50th node, all investigated addresses either represented a small crypto exchange, with transaction volume statistically irrelevant for our analysis, or did not behave like a crypto exchange. 

For each of them, we verified whether the corresponding Polkadot address was known to the Polkadot community by relying on off-chain data. More specifically, we found the identity of 17 exchanges in the Subscan\footnote{\url{https://polkadot.subscan.io/}} blockchain explorer---data from Merkle Science\footnote{\url{https://www.merklescience.com/}}. The remaining 16 are unknown to the Polkadot community, and we found no information about the identity of the exchange. Therefore, we labeled these exchanges as ``unknown'', followed by a hexadecimal id to distinguish them. The top 10 exchanges are reported in Table~\ref{tab:exchanges}, sorted by the number of wallet addresses. 
The first column reports the name, the second one the number of main wallet addresses, while the third column shows the overall number of owned Polkadot addresses, including main wallet and deposit addresses. The fourth column, ``Inter-exchange transactions'', refers to transactions directed to or coming from other exchanges, and will be discussed in Section~\ref{sec:results}.

\begin{table}
    \renewcommand{\arraystretch}{1.3}
    \centering
    \begin{tabular}{c|c||c|c|c}
        \textbf{Exchange} & \textbf{No. main} & \textbf{Number of} & \multicolumn{2}{c}{\textbf{Inter-exchange transactions}} \\
        \textbf{name} & \textbf{addresses} & \textbf{nodes} & \textbf{number} & \textbf{amount} \\
        \hline
        \hline
        \multirow{2}{*}{Binance} & \multirow{2}{*}{2} & 195678 & 321722 & 620116608.0 \\
         & & (22.29\%) & (19.16\%) & (37.32\%) \\
        \hline
        \multirow{2}{*}{Kraken.com} & \multirow{2}{*}{1} & 122263 & 311118 & 456932096.0 \\
         & & (13.93\%) & (18.53\%) & (27.50\%) \\
        \hline
        \multirow{2}{*}{Coinbase.com} & \multirow{2}{*}{8} & 115178 & 269823 & 92455408.0 \\
         & & (13.12\%) & (16.07\%) & (5.56\%) \\
        \hline
        \multirow{2}{*}{Kucoin.com} & \multirow{2}{*}{1} & 86065 & 146344 & 23823488.0 \\
         & & (9.80\%) & (8.71\%) & (1.43\%) \\
        \hline
        \multirow{2}{*}{\textbf{Unknown 1}} & \multirow{2}{*}{1} & 66044 & 138718 & 12135886.0 \\
         & & (7.52\%) & (8.26\%) & (0.73\%) \\
        \hline
        \multirow{2}{*}{Okx.com} & \multirow{2}{*}{1} & 40789 & 70031 & 92998136.0 \\
         & & (4.65\%) & (4.17\%) & (5.60\%) \\
        \hline
        \multirow{2}{*}{Huobi.com} & \multirow{2}{*}{1} & 35672 & 66618 & 168555936.0 \\
         & & (4.06\%) & (3.97\%) & (10.14\%) \\
        \hline
        \multirow{2}{*}{ChangeNow.io} & \multirow{2}{*}{1} & 31645 & 31654 & 1819636.625 \\
         & & (3.60\%) & (1.88\%) & (0.11\%) \\
        \hline
        \multirow{2}{*}{Nexo} & \multirow{2}{*}{1} & 30681 & 49471 & 8417546.0 \\
         & & (3.49\%) & (2.95\%) & (0.51\%) \\
        \hline
        \multirow{2}{*}{HitBTC} & \multirow{2}{*}{1} & 25420 & 26328 & 2389175.25 \\
         & & (2.90\%) & (1.57\%) & (0.14\%)
    \end{tabular}
    \caption{Top 10 largest exchange services (w.r.t.\ the number of nodes) with corresponding statistics}
    \label{tab:exchanges}
\end{table}

\subsection{Contraction of the Network}
\label{subsec:contraction-setup}

With the insight that more than 800k accounts belong to 33 exchange services only, we decided to contract the graph in order to analyze their impact on the overall network.

We labeled all the exchange accounts (alongside with the corresponding deposit addresses) belonging to one of the $33$ different exchange services with a different color $c_i=i$ with ${i=1, \dots, 33}$.
Then, we set a special color, say $c=0$, for all the unclassified remaining accounts. In what follows, we refer to those accounts as end-users (or simply \emph{users}) for the sake of readability.
Nevertheless, it is worth noticing that such nodes contain both regular users, small exchanges we have not detected, system users, e.g., validators, nominators, proxy nodes, and, potentially, other roles that are still unveiled.

Hence, we have a coloring function:
\begin{equation}
    \label{eq:gamma}
    \begin{matrix}
        \gamma &:& V &\to& \{0, \dots, 33\}\\
         & & v & \mapsto & c
    \end{matrix}
\end{equation}

Evaluating $G/\gamma$ as described in Section~\ref{subsec:contraction} unleashes three different useful capabilities:
\begin{enumerate}
    \item[(i)] The network gets shrank in size while retaining the main connectivity information: this provides a computational speed-up without a significant loss in information and, as a side effect, navigating it also gets simpler.
    
    \item[(ii)] All deposit addresses are absorbed into their corresponding main address, implicitly simplifying exchanges structure and highlighting their impact on the network (in terms of coverage, connectivity, and flux moved).

    \item[(iii)] User accounts are inherently divided into clusters depending on their mutual interaction, where a single node is created in place of each set of user nodes interacting together. This allows a fast and simple detection of all users who interact exclusively with exchange services.
\end{enumerate}

In the following figures, meant to capture the idiosyncratic features highlighted by our analysis, we will depict exchange-related and user-related statistics in blue and orange respectively.

\medskip

From an operational point of view, in order to solve the task of evaluating the contraction efficiently, we adopted the greedy iterative approach introduced in~\cite{LombardiOnofri22a} and later extended in~\cite{LombardiOnofri22b}.
In particular, for what concerns the results discussed later in Section~\ref{sec:results}, we modified the online available\footnote{\url{https://github.com/eOnofri04/GraphColorContraction}} C-lang contraction framework to tackle the cryptocurrency network;
in fact, it was originally designed as a minimal working tool to prove the effectiveness of the methodology on simple undirected graphs and required the proper improvements to be used in practice.

The main characteristic of the transaction graph lies in the value of \emph{DOT}s each transaction (\ie the edges $\bar E$ of the multigraph) moves from the sender to the recipient.
Given that the contraction is a procedure that aims at reducing the total size of the graph, it is then natural to aggregate all the transactions occurring from a given sender to a given recipient in a single edge, hence forming a simple di-graph $G = (V, E)$ sharing the vertices with $\bar G$ and where a single edge carries the aggregated information of multiple transactions.
This is the reason why we equipped each edge $e = (u, v) \in E$ with a floating point value representing the total amount of \emph{DOT}s moved from $u$ to $v$ and an integer counter of the number of transactions the edge represents.
In what follows, we will refer to these values as \emph{flux} and \emph{multiplicity} of the edge.

This simple extension naturally provides a way to apply the framework to di-graphs as well. 
In fact, whenever $(u, v) \in E$ but $(v, u) \not\in E$, a novel edge with multiplicity 0 and undefined flux can be added.

It is natural to retain flux and edge multiplicity that moves within clusters as well.
Hence, we also equipped all the vertices $v \in V$ with three novel statistics: a floating point value retaining the flux circulating within the cluster and two integer counters accounting the number of the intra-cluster transactions and the number of aggregated nodes.

Finally, we also introduced a way to explicitly track the cluster $v' \in V'$ each node $v \in V$ belongs to.

\section{Results and Discussion}\label{sec:results}

In this section, we introduce the results obtained applying the $\gamma$-contraction to the transaction graph $G = (V, E)$ given the coloring for $\gamma$ provided by the exchange clustering step, as in \eqref{eq:gamma}.
To ease the reading, we recall that the order of $G$ is  $n = 2.3 \times10^6$ nodes, its size accounts for $m = 8.1 \times 10^6$ edges, and the total flux being $f = 9.2e+9$ DOTs (see Table~\ref{tab:networkStatistics}).

Our modified contraction algorithm converged to $G/\gamma$ in three iterations, taking less than 20 minutes on a Ryzen 5 5600X 6 core 32Gb Ram desktop computer running Ubuntu 20.04 LTS. As a result, we obtained a contracted graph $G' = (V', E') = G/\gamma$ of order $n' = 6.54 \times 10^5 \sim \sfrac{3}{10} \cdot n$ and size $m' = 9.4\times10^5 \sim \sfrac{1}{10} \cdot m$.

\begin{figure}
    \centering
    \begin{subfigure}{.49\linewidth}
        \includegraphics[width=\linewidth]{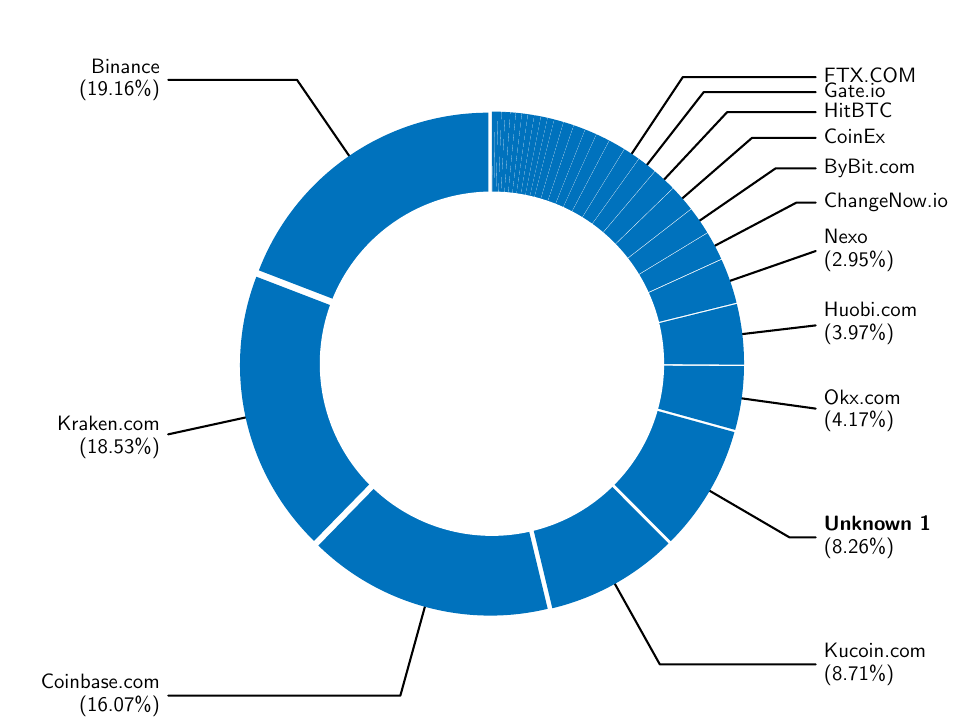}
        \caption{}
        \label{fig:distribution_EE_ec}
    \end{subfigure}
    \begin{subfigure}{.49\linewidth}
        \includegraphics[width=\linewidth]{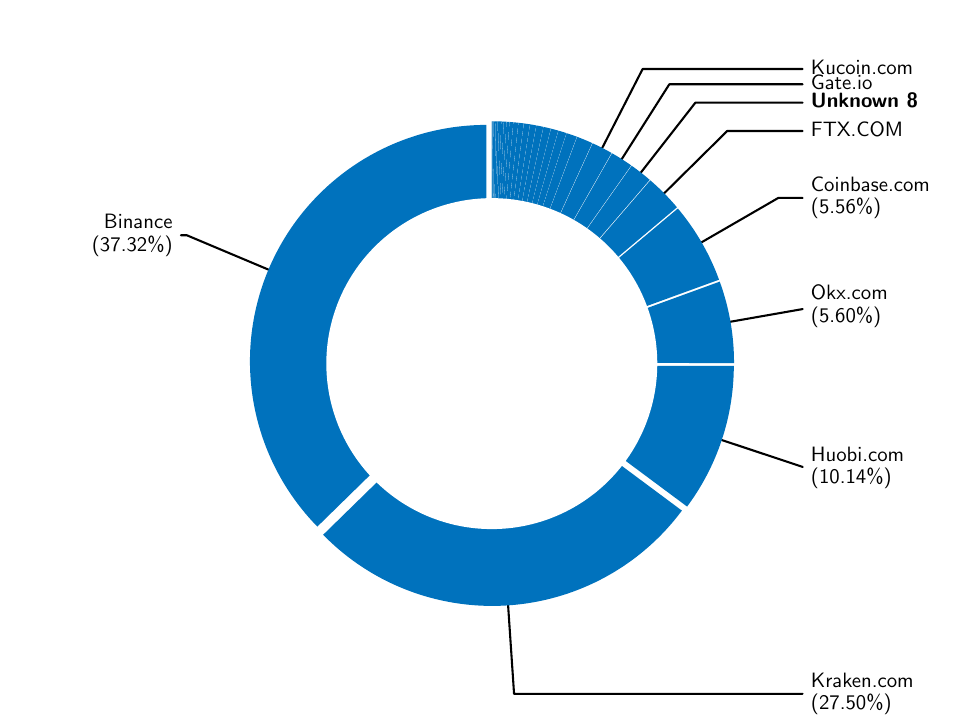}
        \caption{}
        \label{fig:distribution_EE_ef}
    \end{subfigure}
    
    \caption{
        Distribution of the intra-exchange transactions. Slices are proportional to \subref{fig:distribution_ec} the number of transactions and \subref{fig:distribution_ef} the total transaction amount, as reported in Table~\ref{tab:exchanges}.
    }
    \label{fig:distribution_EE}
\end{figure}

\begin{figure}\centering
    \def\dx{.49}
    \begin{subfigure}{\dx\linewidth}
        \includegraphics[width=.95\linewidth]{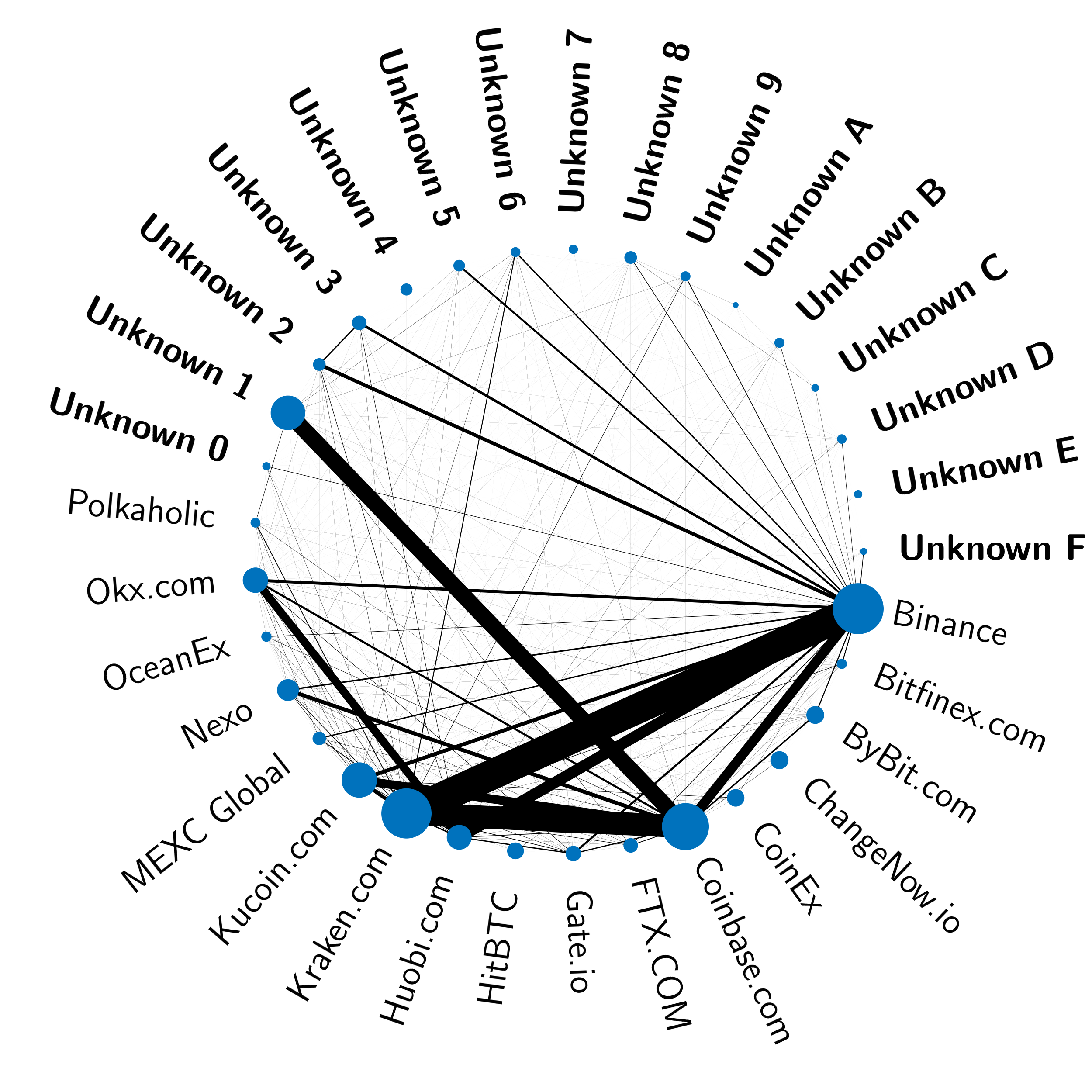}
        \caption{}
        \label{fig:graph_ec_EE}
    \end{subfigure}
    \begin{subfigure}{\dx\linewidth}
        \includegraphics[width=.95\linewidth]{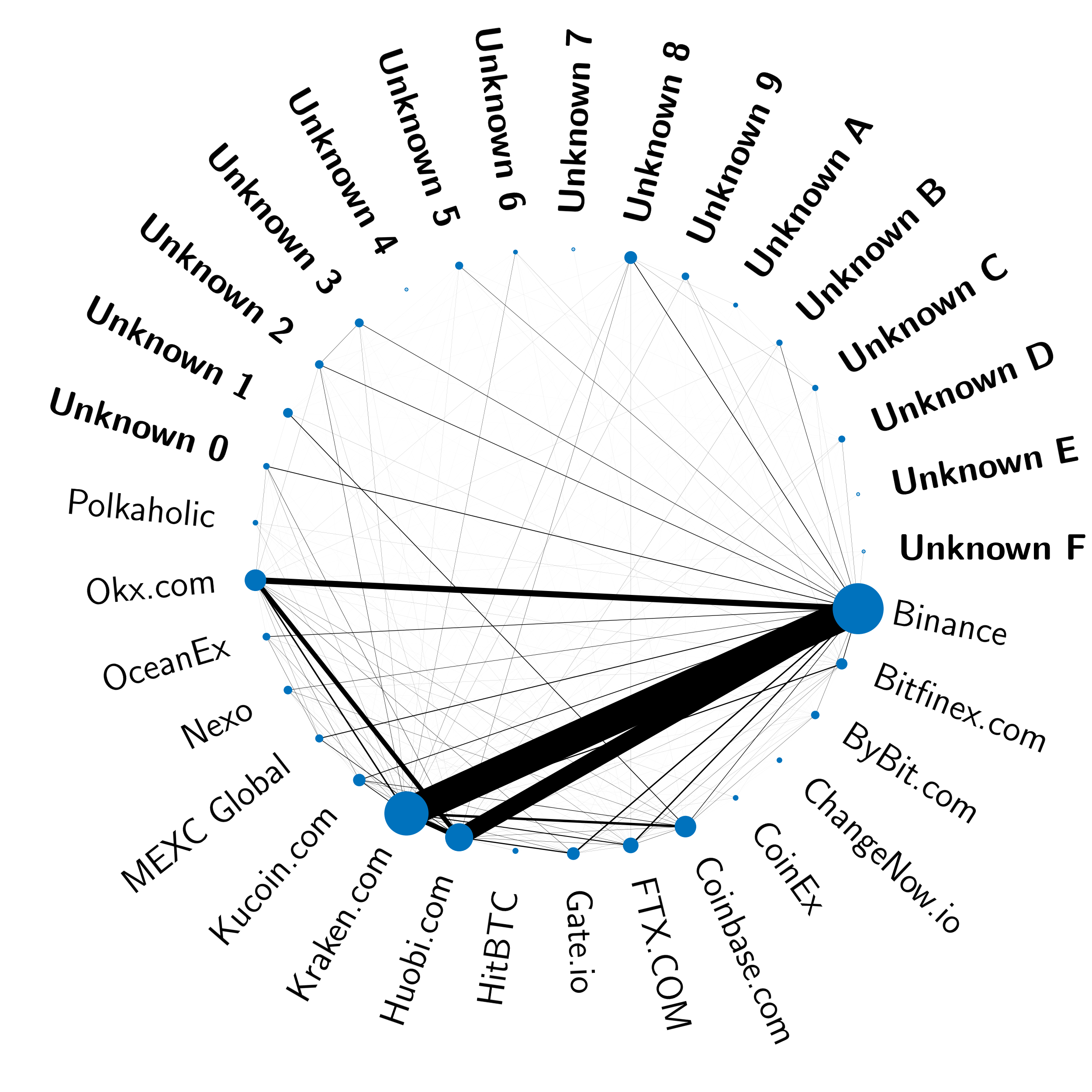}
        \caption{}
        \label{fig:graph_ef_EE}
    \end{subfigure}
    \caption{
        Exchanges ecosystem within the contracted graph represented in terms of \subref{fig:graph_ec_EE} number of transactions and \subref{fig:graph_ef_EE} total transaction amount.
        Nodes size is proportional to the intra-cluster transactions reported in Table~\ref{tab:exchanges} while edges width is proportional to the related inter-exchange set of transactions.}
    \label{fig:graph_EE}
\end{figure}

\begin{figure}
    \centering
    \def\dx{.49}
    \begin{subfigure}{\dx\linewidth}
        \includegraphics[width=\linewidth]{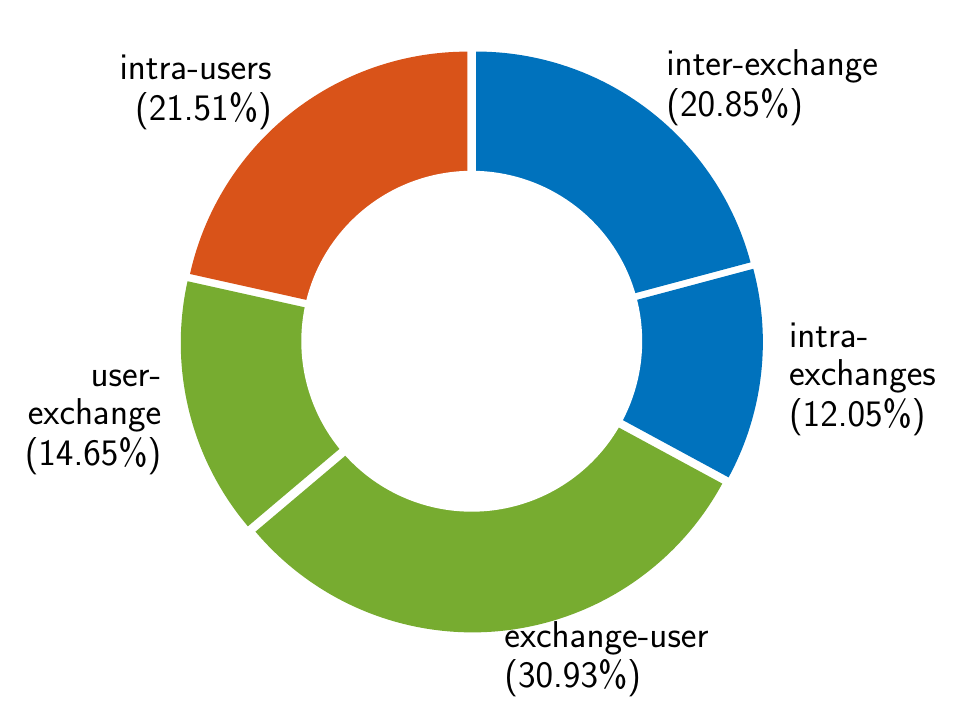}
        \caption{}
        \label{fig:distribution_ec}
    \end{subfigure}
    \begin{subfigure}{\dx\linewidth}
        \includegraphics[width=\linewidth]{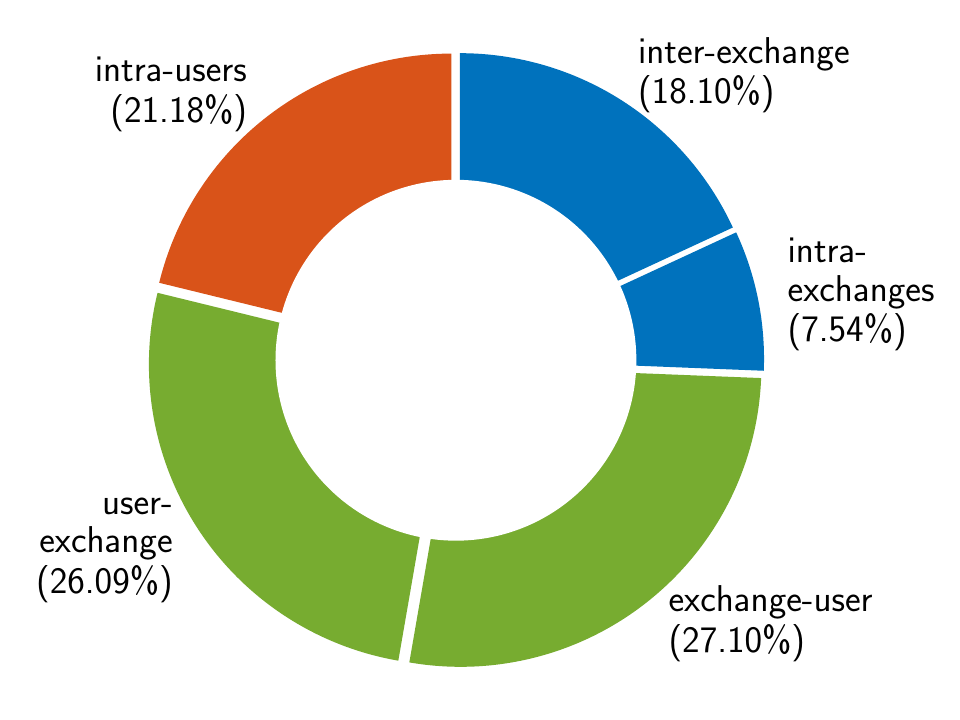}
        \caption{}
        \label{fig:distribution_ef}
    \end{subfigure}
    
    \caption{
        Distribution of the transactions within the network in terms of \subref{fig:distribution_ec} number and \subref{fig:distribution_ef} total amount.
    }
    \label{fig:distribution_e}
\end{figure}

All 33 exchange services converge to a single node each, thus forming 33 nodes of $G'$.
This fact is not surprising for exchanges with a single main wallet, like \textit{Kraken.com} and \textit{Kucoin.com}. However, it highlights that exchange ecosystems with multiple central nodes, such as \textit{Binance} and \textit{CoinBase.com}, are strictly entangled, with money transfers occurring within the different central nodes.
It is worth noting that the total  flux for movements within a very same exchange spans $12.05\%$ of the total flux of the network, corresponding to $7.54\%$ of the total value, as shown in Figure~\ref{fig:distribution_e}.
In order to provide more details on such nodes, the statistics associated with the transactions occurring within the ten largest exchange services are summarized in the fourth column of Table~\ref{tab:exchanges}, in terms of absolute and relative values.
A pictorial representation of these data is also reported in Figure~\ref{fig:distribution_EE}, where the intra-exchange transaction distribution is qualitatively provided for the reader's convenience.
While it is not surprising that exchanges with several main addresses 
like \emph{CoinBase.com} issue a large number of internal transactions, it is worth noting that nearly $\sfrac23$ of the internal flux is ascribable only to \textit{Binance} and \textit{Kraken.com}. This can be seen as a sign of the overwhelming impact of the few largest exchanges active on the Polkadot network, dominating the trading market.
Although justifiable with the natural money flow from deposit addresses to the exchange main wallet, a high intra-exchange transaction volume might suggest wash trading activities, and should be carefully analyzed in future work.

Even more noticeable is the fact that most of the exchanges are directly inter-connected via multiple edges, as depicted in Figure~\ref{fig:graph_EE}, which shows the subgraph spanned by the $33$ exchange nodes. 
Following this trend, it is worth noting that a considerable $18.10\%$ of the total transaction flux, corresponding to $20.85\%$ of the total number of transactions, is moved via inter-exchanges transactions (see Figure~\ref{fig:distribution_e}).
From Figure~\ref{fig:graph_EE}, we can observe different ``communities'' of exchanges, with \emph{Binance} standing out as the main player, interacting with almost all other exchanges, and encompassing the vast majority of traffic, in terms of both volume and flux. \emph{Coinbase.com}, instead, appears very active in terms of number of transactions, particularly with \emph{Kraken.com} and \emph{Unknown 1}. However, as shown in Figure~\ref{fig:graph_ef_EE}, the moved amount of \emph{DOT}s is very little compared to the flux moved by the \emph{Binance} ecosystem.
These inter-exchange transactions might be due to trading purposes, with end-users moving \emph{DOT}s from one account on an exchange platform to a different account on a second exchange service. However, these transactions might also be due to direct interactions between exchanges for various purposes, such as money lending or other kind of collaborations, which we plan to investigate in details in future research.

Figure~\ref{fig:distribution_e} also highlights one of the most interesting results of our analysis: more than a half ($53.10\%$) of the total flux is moved from users to exchanges and vice versa. This lays the foundation of our conjecture that the Polkadot cryptocurrency has been used, so far, mainly for speculative investment; in fact, way less than a fourth of the flux belongs to user-to-user transactions (recall that within the $21\%$ intra-users slice also lie different roles other than end-users, as discussed in the beginning of Section~\ref{subsec:contraction-setup}).

\begin{figure}
    \includegraphics[width=\linewidth]{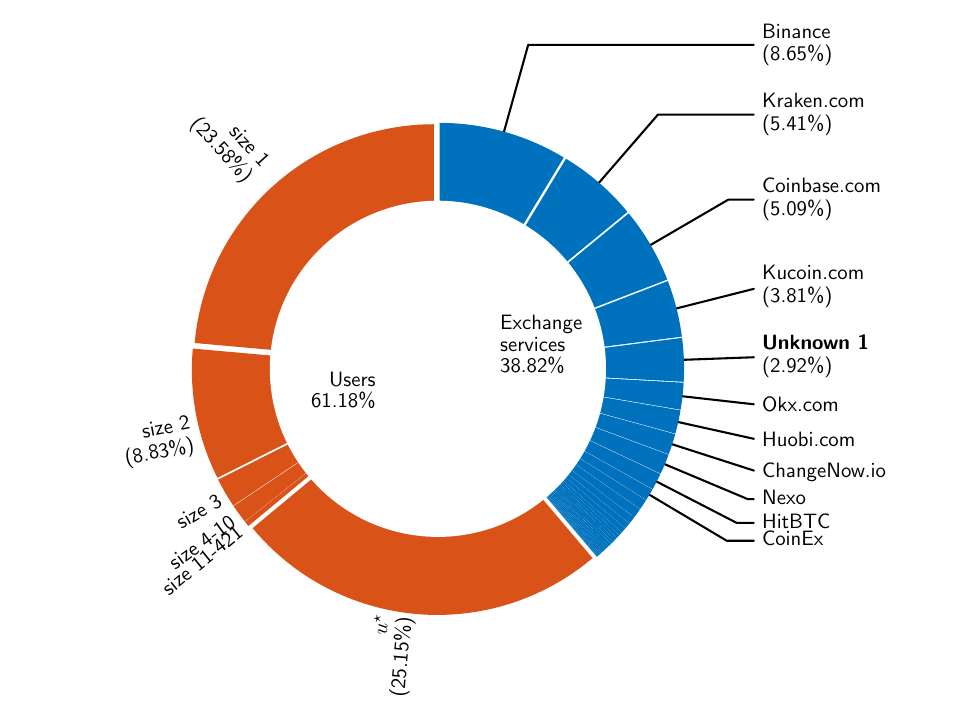}
    \caption{
        Distribution of the nodes in the original network $G$.
        Exchanges are segmented according to the number of nodes each exchange service is made of, as detailed in Table~\ref{tab:exchanges}.
        Users are segmented depending on the size of the cluster they belong to, as detailed in Table~\ref{tab:nodes_distribution}.
        Do note that users in `size 1' are isolated w.r.t.\ other users.
    }
    \label{fig:nodes_distribution}
\end{figure}
\begin{figure}
    \centering
    \includegraphics[width=0.98\linewidth]{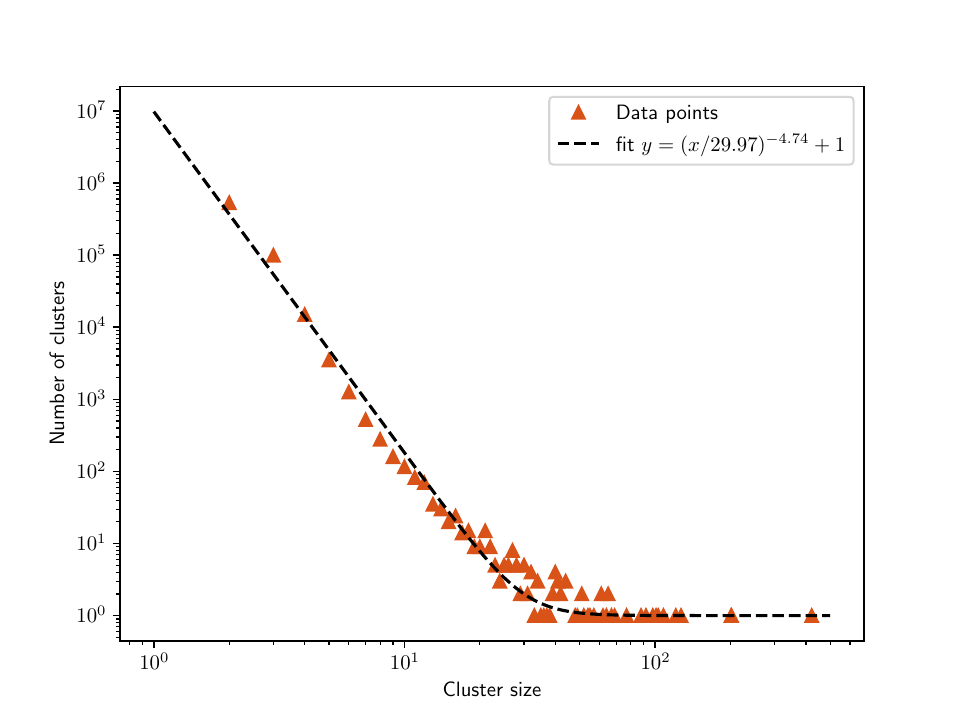}
    \caption{
        Distribution of the user clusters with respect to their size (see also Table~\ref{tab:nodes_distribution}).
        The black dashed line highlights the power-law behaviour.
    }
    \label{fig:UU_avc_distribution}
\end{figure}
\begin{table}
    \renewcommand{\arraystretch}{1.3}
    \centering
    \scriptsize
    \begin{tabular}{c||c|c|c|c|c|c|c|c}
        \textbf{Size} & 1 & 2 & 3 & 4--10 & 11--100 & 101--421 & 422--568822 & 568823 \\
        \hline\hline
        \multirow{2}{*}{\textbf{Number of clusters}} & \multirow{2}{*}{533308} & \multirow{2}{*}{99824} & \multirow{2}{*}{15006} & \multirow{2}{*}{5930} & \multirow{2}{*}{337} & \multirow{2}{*}{7} & \multirow{2}{*}{0} & \multirow{2}{*}{1}\\
        &&&&&&&&\\\hline
        \multirow{2}{*}{\textbf{Number of users}} & 533308 & 199648 & 45018 & 28573 & 6875 & 1277 & 0 & 568823\\
        & (38.55\%) & (14.43\%) & (3.25\%) & (2.07\%) & (0.50\%) & (0.09\%) & (0.0\%) & (41.11\%) \\\hline
        \textbf{Average intra-cluster} & \multirow{2}{*}{0} & \multirow{2}{*}{1.37} & \multirow{2}{*}{3.18} & \multirow{2}{*}{7.24} & \multirow{2}{*}{52.43} & \multirow{2}{*}{284.85} & \multirow{2}{*}{--} & \multirow{2}{*}{1485582.0}\\
        \textbf{transactions number} &&&&&&&\\\hline
        \textbf{Average intra-cluster} & \multirow{2}{*}{0} & \multirow{2}{*}{153.69} & \multirow{2}{*}{878.23} & \multirow{2}{*}{8672.67} & \multirow{2}{*}{39365.29} & \multirow{2}{*}{4403.12} & \multirow{2}{*}{--} & \multirow{2}{*}{1851260928.0}\\
        \textbf{transactions flux} &&&&&&&\\
    \end{tabular}
    \caption{Users distribution (w.r.t.\ cluster sizes) with corresponding statistics}
    \label{tab:nodes_distribution}
\end{table}

Further evidence suggesting a speculative behavior can be extrapolated from Figure~\ref{fig:nodes_distribution}, reporting the distribution of the accounts in the original network w.r.t.\ their role.
In particular, two main evidence arise: (i) at least $\approx \sfrac25$ of the $n$ nodes of the network belong to the 33 exchanges services detected; and, (ii) only $\approx \sfrac14$ of the $n$ nodes belongs to potential end users that perform transactions one with the others. 
This last set of nodes is represented as a single node $u^\star$ in $G'$, accounting for $568,823$ nodes from the original network $G$ ($41.11\%$ of the users). 
Despite we classified them as end users, the remaining $35\%$ of the nodes interact with no or little other nodes but exchanges---this fact suggesting that our conservative estimates of the role of exchanges are indeed a lower bound. 
The exact numbers and sizes of users clusters from $G'$ are reported in Table~\ref{tab:nodes_distribution} and plotted in Figure~\ref{fig:UU_avc_distribution} to highlight their power-law tendency, where we recall that a cluster of $x$ users in $G'$ corresponds to $x$ users transacting with $x-1$ other users in $G$ (cf.\ users segments in Figure~\ref{fig:nodes_distribution}).

\begin{figure}
    \centering
    \def\dx{.49}
    \begin{subfigure}{\dx\linewidth}
        \includegraphics[width=\linewidth]{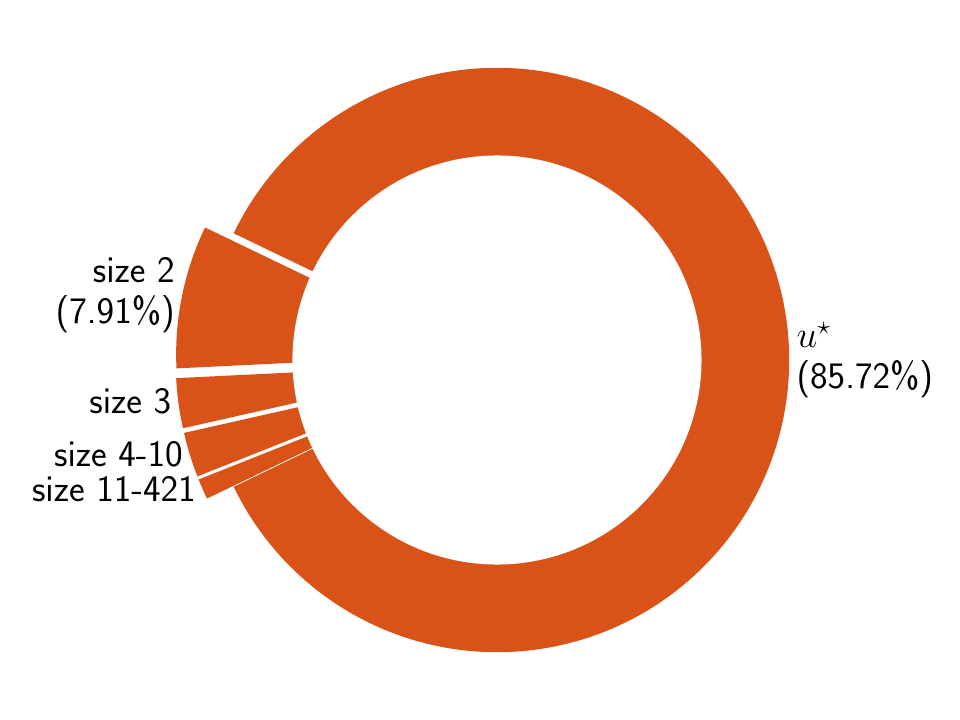}
        \caption{}
        \label{fig:distribution_UU_ec}
    \end{subfigure}
    \begin{subfigure}{\dx\linewidth}
        \includegraphics[width=\linewidth]{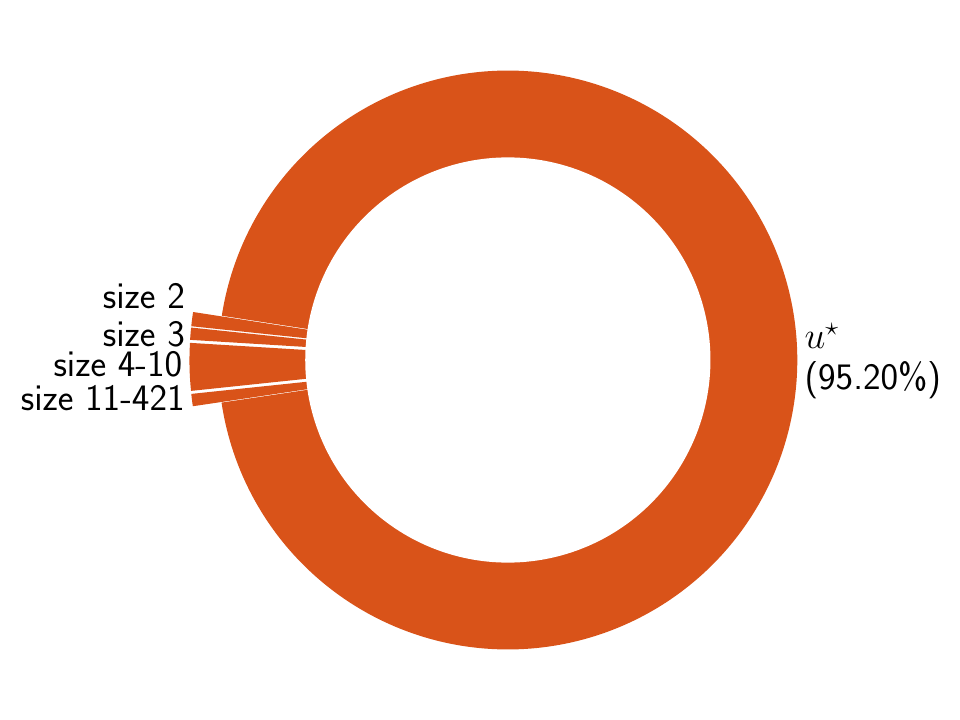}
        \caption{}
        \label{fig:distribution_UU_ef}
    \end{subfigure}
    
    \caption{
        Distribution of the intra-cluster transactions amongst the users in terms of \subref{fig:distribution_UU_ec} number of transactions and \subref{fig:distribution_UU_ef} total transaction amount, partitioned depending on the size of the cluster they belong to, as detailed in Table~\ref{tab:nodes_distribution}.
    }
    \label{fig:distribution_UU}
\end{figure}

From a practical point of view, this means that $38.55\%$ of the users (cf.\ Table~\ref{tab:nodes_distribution}) never transacted with any other user. We believe this is a good indicator to classify such users as traders, with little to no impact on the cryptocurrency ecosystem---at least as per the objectives the cryptocurrency is meant to support.
A similar outcome can be inferred for most of the remaining $20.33\%$ user nodes not collapsing into $u^\star$; in fact, as it can be seen from Figure~\ref{fig:distribution_UU}, less than $5\%$ of the total flux can be ascribed to those users.

\begin{figure}\centering
    \def\dx{.49}
    \begin{subfigure}{\dx\linewidth}
        \includegraphics[width=\linewidth]{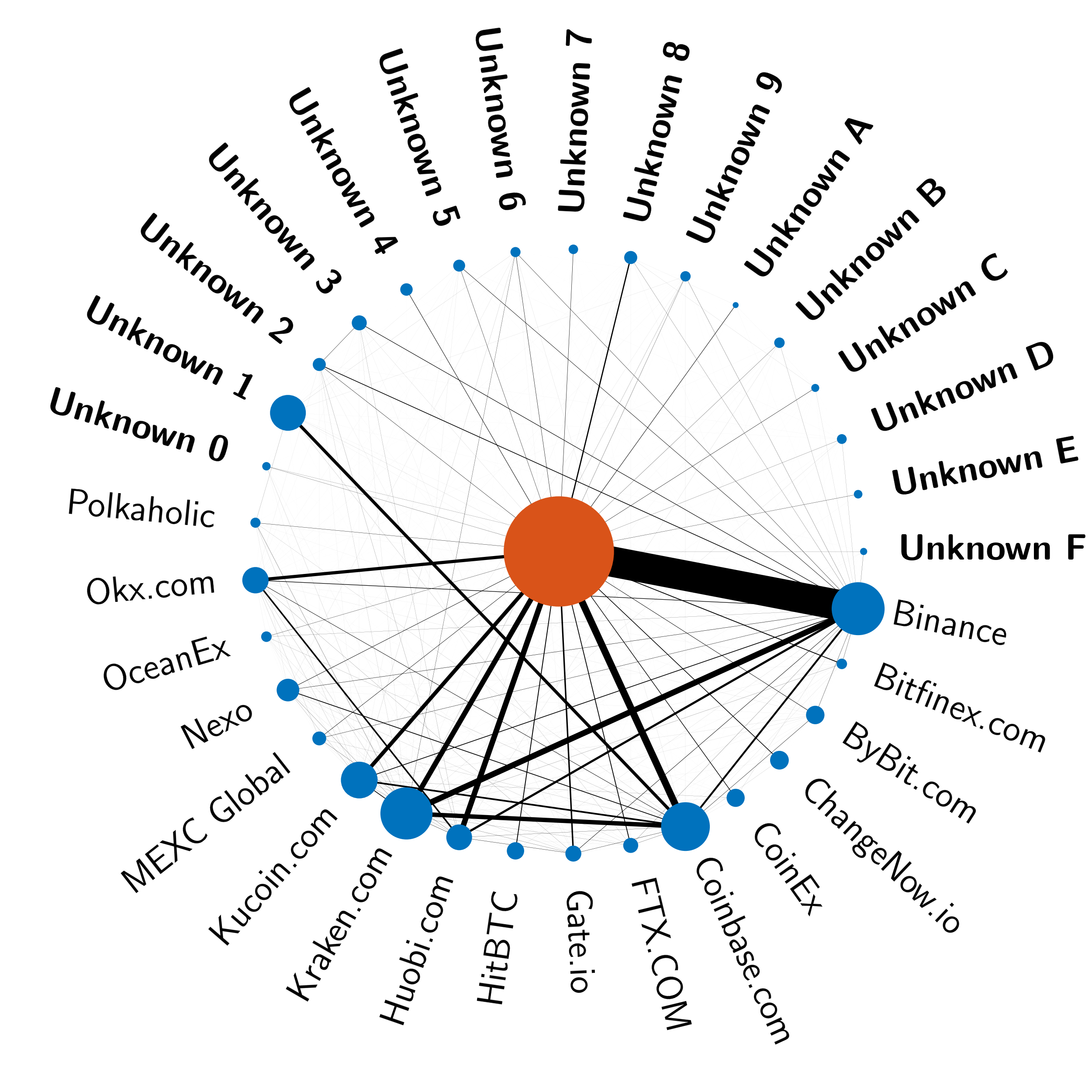}
        \caption{}
        \label{fig:graph_ec_UE}
    \end{subfigure}
    \begin{subfigure}{\dx\linewidth}
        \includegraphics[width=\linewidth]{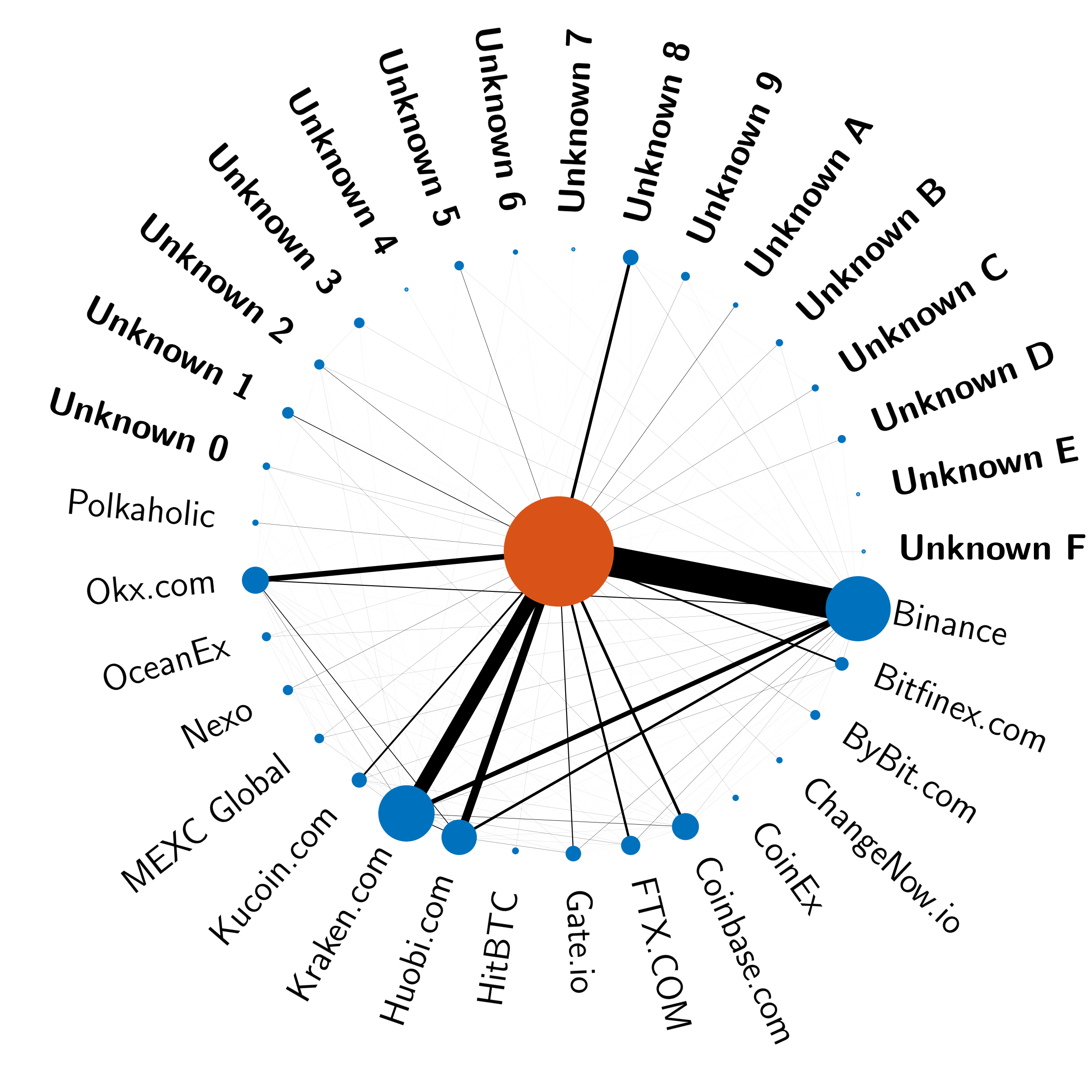}
        \caption{}
        \label{fig:graph_ef_UE}
    \end{subfigure}
    \caption{
        Effective ecosystem within the contracted graph, refinement of the exchanges ecosystem from Figure~\ref{fig:graph_EE} in terms of
        \subref{fig:graph_ec_UE} number of transactions and
        \subref{fig:graph_ef_UE} total transaction amount.
    }
    \label{fig:graph_UE}
\end{figure}

From the cited data, we can prove an upper bound on the number of effective users to $\sim 0.5 \times 10^6$, all collapsing to $u^\star$ in $G'$, \ie $\approx \sfrac14$ of the original network.

A quantitative representation of the cluster $u^\star$ with respect to the 33 exchanges (along with the mutual interactions) can be found in Figure~\ref{fig:graph_UE}, that represents the subgraph of $G'$ spanned by those $34$ vertices.
This once again confirms the predominance of \textit{Binance} and \textit{Kraken.com}, followed by \textit{Huobi.com}, as the top three exchange services influencing the ecosystem. 
This information suggests the idea that Polkadot is currently represented by a few large actors that can potentially heavily affect the future behavior of this project, especially because some of them, such as \emph{Binance}, are actively involved in the protocol as nominator/validator~\cite{abbas2022analysis}.

\section{Conclusion}\label{sec:conclusion}
In this paper, we introduced a novel methodology for analyzing large cryptocurrency transaction graphs to gain insights on the  interactions among different user categories. 
Our solution intends to propose a robust methodology to assess the healthiness of a crypto project while not being subject to the exposed limitations of other commonly used metrics (transactions and related volumes).
Our approach involves constructing the transaction graph, clustering user categories of interest, and contracting the graph to enhance the visualization of interactions among clusters. We applied this methodology to a use case developed over Polkadot. In particular, we studied the Polkadot transaction graph and derived a compacted representation of it after identifying 41 accounts belonging to 33 crypto-exchanges.
Our objective is to shed light on the possible exposure of the Polkadot project to speculative operations. 

Our findings reveal that exchanges exert considerable influence over the Polkadot network, owning nearly 40\% of all addresses in the ledger, including both main wallets and deposit addresses. Moreover, the ecosystem shaped by their presence, which encompasses exchanges themselves and end-users who exclusively interact with exchanges, accounts for almost 80\% of all transactions in the Polkadot ledger. Consequently, the remaining portion of the transaction network, which should include transactions related to other use cases beyond inter-/intra- exchanges transactions and speculation, is minimal. Only a mere 20\% of all transactions in the network do not involve the major detected exchanges. Moreover,  the cited 20\% is an upper bound, since this cluster might contain small exchanges we did not identify, protocol participants like nominators and validators, parachains, and possibly other yet-unveiled categories. This indicates that the proportion of transactions associated with end-user-to-end-user activities is even smaller, suggesting a predominant utilization of the DOT coin for speculative purposes.

Our methodology is general and adaptable. Indeed, it can be applied to other account-based cryptocurrencies as is and it can be easily extended to work with UTXO-based projects and support different categories of users---an activity we are currently working on. 
Finally, we believe that the sound rationale, the detailed methodology, and the extensive experimental results shown in this paper demonstrate the quality and viability of our approach with respect to the adoption of competing metrics that intend to proxy the quality of crypto projects.

\SkipTocEntry
\section*{Acknowledgments}
\begingroup\small
    Flavio Lombardi and Elia Onofri are also members of the ``Gruppo Nazionale Calcolo Scientifico-Istituto Nazionale di Alta Matematica'' (GNCS-INdAM).
\endgroup
\balance
\bibliographystyle{IEEEtran}

\bibliography{ref}

\end{document}